\documentclass[12pt,a4paper]{article}
\usepackage[hmargin=1.5cm,vmargin=1.5cm]{geometry}

\usepackage{lipsum}
\usepackage{authblk}
\usepackage{amssymb}
\usepackage{amsmath}
\usepackage{gensymb}
\usepackage{todonotes}
\usepackage{tabularx}
\usepackage{multirow}
\usepackage{bm}
\usepackage{dblfloatfix}
\usepackage{graphicx}
\usepackage{hyphenat}
\usepackage{colortbl}
\usepackage{cite}
\usepackage{pdfpages}
\usepackage{subcaption}
\usepackage[english]{babel}
\usepackage{url}
\usepackage{fancyhdr}
\usepackage{bibentry}
\usepackage[breaklinks,colorlinks = true,linkcolor = blue,urlcolor=cyan,citecolor=blue]{hyperref}

\def\rim#1{\textcolor{red}{#1}}

\begin{document}

\title{\textbf{Pressure-induced 1T to 3R structural phase transition in metallic VSe$_2$: X-ray diffraction and first-principles theory}}
\author{Srishti Pal,$^1$* Koyendrila Debnath,$^2$* Satyendra Nath Gupta,$^1$ Luminita Harnagea,$^3$ D.V.S. Muthu,$^1$ Umesh V. Waghmare$^2$ and A.K. Sood$^1$\vspace{0.05cm} \\$^1$\small\emph{Department of Physical Sciences, Indian Institute of Science, Bengaluru - 560012, India}\\ 
$^2$\small\emph{Theoretical Sciences Unit, Jawaharlal Nehru Centre for Advanced Scientific Research, Bengaluru - 560064, India}\\
$^3$\small\emph{Department of Physics, Indian Institute of Science Education and Research, Pune - 411008, India}}
\date{}
\maketitle\thispagestyle{empty}

\begin{abstract}
We study pressure-induced structural evolution of vanadium diselenide (VSe$_2$), a 1T polymorphic member of the transition metal di-chalcogenide (TMD) family using synchrotron-based powder X-ray diffraction (PXRD) and first-principles density functional theory (DFT). Our XRD results reveal anomalies at P $\sim$4 GPa in \textit{c/a} ratio, V-Se bond length and Se-V-Se bond angle signalling an isostructural transition. This is followed by a first order structural transition from 1T (space group \textit{P$\bar{3}$m1}) phase to a 3R (space group \textit{R$\bar{3}$m}) phase at P $\sim$11 GPa due to sliding of adjacent Se-V-Se layers. Both the transitions at $\sim$4 and 11 GPa are cognate with associated changes in the Debye-Waller factors, hitherto not reported so far. We present various scenarios to understand the experimental results within DFT and find that the 1T to 3R transition is captured using spin-polarized calculations with Hubbard correction (\textit{U$_{eff}$} = \textit{U} - \textit{J} = 8 eV) giving transition pressure of $\sim$9 GPa, close to the experimental value.
\end{abstract}

\maketitle

\newpage

\section{\bf{INTRODUCTION}}

Layered quasi-2D transition metal di-chalcogenides (TMD; MX$_2$, M = transition metal Mo, W, V, Ta, Ti, Mn etc., X = chalcogen S, Se, Te etc.) are being pursued intensely in recent years due to their emergent properties and significant applications~\cite{Yuki,Zhang,10,13,1}. One of the interesting features of the bulk MX$_2$ compounds is their crystallization into different polytypes (viz. 1T, 2H, 3R, 1T', T$_d$ etc.) depending on the co-ordination of the nearest-neighbour chalcogen polyhedra around the transition metal and the various stacking sequences of the 2D layers in (001) direction~\cite{22,23,24,25,26}. A TMD monolayer is comprised of a sandwiched hexagonal layer of transition metal atoms between two hexagonal layers of chalcogens. These covalently bonded  X-M-X slabs are then stacked in vertical \textit{c}-direction with weak van der Waals bond between them, resulting in an anisotropic 3D structure. The trigonal 1T polytype with standard CdI$_2$ structure having space group \textit{P$\bar{3}$m1}, the only stable structure of VSe$_2$ in ambient conditions~\cite{27,28,29}, belongs to a regular octahedral co-ordination of six Se atoms around the central vanadium and a stacking sequence \textit{aBc} (where \textit{a, b} and \textit{c} label Se atomic layers and \textit{A, B} and \textit{C} label V atomic layers) of Se-V-Se monolayers without any lateral shift (Fig.~\ref{fig1}(a)). The 3R polytype also has the same regular octahedral configuration but with a lateral shift between three successive Se-V-Se layers (\textit{aBc bCa cAb} stacking) (Fig.~\ref{fig1}(b)).

Unlike the semiconducting 2H polytypes, the 1T bulk VSe$_2$ is metallic due to significant overlaps between vanadium \textit{d} bands and selenium \textit{p} bands and shows a charge density wave (CDW) state~\cite{32,33}. 1T-VSe$_2$ is unique in the formation of a 3D chiral CDW~\cite{28,34,35} due to the partial nesting of its Fermi surface~\cite{37,38}. X-ray and electron diffraction studies~\cite{39,40} have established periodic lattice deformation to be the key precursor to this CDW transition to an incommensurate phase below 110 K and to a \textit{4a'}$\times$\textit{4a'} commensurate superlattice structure below 80 K (still incommensurate along \textit{c}-axis with \textit{c'} $\approx$ \textit{3c}). The CDW transition of 1T-VSe$_2$ is very sensitive to any external perturbation that directly affects its electronic band structure. While the effects of reduced thickness down to monolayer limit~\cite{41,42,43,44,Ganbat} or intercalation by Na, K, Cs etc.~\cite{45,46,47} or interstitial vanadium itself~\cite{48} have been studied, there is limited work on its electronic and structural stability under external pressure. While a large number of TMD crystals like 2H$_c$-MoS$_2$, MoSe$_2$, WSe$_2$ exhibit pressure induced metallization followed by an isostructural transformation of the crystal to 2H$_a$ (except for 2H$_c$-MoSe$_2$)~\cite{49,50,51,52}, a few others show crystal symmetry change, e.g. transformation of trigonal 1T-TiS$_2$ to an orthorhombic phase at 16.2 GPa~\cite{53}, cubic to orthorhombic transition of MnS$_2$~\cite{54}, appearance of monoclinic phase in 1T-IrTe$_2$ at 5 GPa followed by transition to a cubic phase at 20 GPa~\cite{55}.

Friend et al.~\cite{56} have reported high pressure resistivity and Hall studies on bulk 1T-VSe$_2$ up to 3 GPa which showed an increase of CDW transition temperature (\textit{T$_C$}) with $\frac{dT_C}{dP} \sim$0.8 K.GPa$^{-1}$ due to pressure broadening of vanadium d-conduction band. Recently, using the crystals from same batch as used in the present study, Sahoo et al.~\cite{57} showed pressure-enhancement of the CDW \textit{T$_C$} in bulk 1T-VSe$_2$ reaching 240 K at 12 GPa followed by quenching of the CDW state before finally evolving into a superconducting phase with \textit{T$_C$} $\sim$4 K after 15 GPa. While this manuscript was under preparation, a recent report on high pressure XRD and DFT studies~\cite{58} indicated the new phase of VSe$_2$ after 12 GPa to be a 3$\times$3$\times$1 supercell of monoclinic symmetry. Here we report high pressure XRD and density functional theoretical (DFT) calculations which are at variance with the recent report~\cite{58}. Our XRD data shows a first order structural transition to a 3R phase at $\sim$11 GPa. We also show an isostructural transition at $\sim$4 GPa, in agreement with recent reports~\cite{57,58}, not only by the anomalous \textit{c/a} ratio but also by the changes in V-Se bond length and Se-V-Se bond angle. Our XRD data and theoretical calculations do not support monoclinic superstructure above 11 GPa. Our detailed DFT calculations do show a transition from 1T to 3R at $\sim$9 GPa, a pressure value very close to the experimental transition pressure of 11 GPa. Notably, the thermal Debye Waller factors of selenium atoms increase by a factor of $\sim$4 across the structural phase transition.

\section{\bf{EXPERIMETAL AND COMPUTATIONAL DETAILS}}\vspace{0.2cm}

Single crystals of VSe$_2$ were grown using a chemical vapor transport technique using iodine as transporting agent. The crystal structure, purity, crystal quality, morphology and chemical composition of the single crystals were determined using X-ray powder diffraction (Bruker D8 diffractometer: Cu-K${_{\alpha}}$ radiation), Laue diffraction and scanning electron microscope (ZEISS GeminiSEM 500) equipped with an energy dispersive X-ray spectroscopy probe (EDX). The samples proved to be homogeneous, with V:Se ratio of 1:2 within the error bar of the techniques (1$-$2 at. $\%$). The electrical characterization of the crystals is reported in our recent work~\cite{57}.

Single crystals of 1T-VSe$_2$ were powdered and loaded inside a Mao-Bell type diamond anvil cell (DAC). The DAC had two 16-facet brilliant cut diamonds with $\sim$600 $\mu$m culet diameter to pressurize the sample placed inside the stainless-steel gasket hole of $\sim$200 $\mu$m diameter. Ruby fluorescence was used to calibrate the applied pressure~\cite{Mao1986}. 4:1 methanol-ethanol mixture was used to transmit the pressure through the sample.

The angle dispersive synchrotron XRD study on 1T-VSe$_2$ has been carried out at Elettra, Italy using the Xpress beamline having $\lambda$ = 0.4957 {\AA} from 0.2 GPa to 26 GPa at room temperature. Data was recorded using MAR 345 image plate. Standard LaB$_6$ crystal was used to calibrate sample to detector distance and orientation angles of the detector. The selected area 2D diffraction pattern was processed using Fit2D software~\cite{Hammersley1998} for conversion into 1D 2$\theta$ vs intensity plot. The raw data for the entire pressure range of 0.2 to 26 GPa was fitted using standard Rietveld refinement procedure in GSAS software package~\cite{Larson2000}.

Our first-principles calculations are based on DFT as implemented in Quantum ESPRESSO (QE) package~\cite{giannozzi2009quantum}, in which we treat only the valence electrons by replacing the potential of ionic cores with pseudopotentials. The exchange-correlation energy of electrons is treated within a generalized gradient approximation (GGA)~\cite{hua1997generalized} with a functional form parametrized by Perdew, Burke, and Ernzerhof~\cite{perdew1998perdew}. Electronic wave functions and charge density were represented in plane wave basis sets truncated at energy cut-offs of 60 Ry and 500 Ry respectively. Brillouin zone (BZ) integrations were sampled on uniform dense $24\times24\times12$ and $24\times24\times6$ meshes of \textbf{k}-points for 1T and 3R structures of VSe\textsubscript{2}, respectively. The discontinuity in occupation numbers of electronic states was smeared using a Fermi-Dirac distribution function with broadening temperature of \textit{k}\textsubscript{B}T = 0.003 Ry. We include van der Waals (vdW) interaction using PBE + D2 method of Grimme~\cite{grimme2004accurate}. Dynamical matrices were calculated within the Density Functional Perturbation Theory (DFPT)~\cite{baroni2001phonons} on a $2\times2\times2$ mesh in the Brillouin Zone. We Fourier interpolated these dynamical matrices to obtain the phonon frequencies at arbitrary wavevectors and dispersion along the high symmetry lines in the Brillouin zone. We also performed  first-principles calculations using projected augmented wave (PAW) method~\cite{blochl1994projector, PhysRevB.59.1758} as implemented in Vienna \textit{ab-nitio} simulation package VASP~\cite{kresse1996efficient, kresse1996efficiency}. Spin-polarized calculations were performed using Perdew-Burke-Ernzerhof (PBE) functional for exchange-correlation term, with Hubbard parameter correction GGA+\textit{U} introduced by Dudarev et al.~\cite{Dudarev,Lutfalla} in which the parameters \textit{U} and \textit{J} do not enter separately, only the difference (\textit{U} –- \textit{J}) is relevant (\textit{U}\textsubscript{eff} = \textit{U - J} = 8.0 eV). Plane wave basis was truncated at a kinetic energy cut-off of 36.75 Ry. Maximum Force on ions in relaxed structure was within a threshold of 10\textsuperscript{-4} eV/{\AA}. A sparse $8\times8\times3$ mesh of k-points was used in sampling Brillouin zone (BZ) integrations in calculations of 1T and \textit{C2/m} structures, similar to a recent report ~\cite{58} and $8\times8\times3$ mesh was used in calculation of 3R structure (primitive unit cell considered).

\section{\bf{RESULTS AND DISCUSSIONS}}\vspace{0.2cm}

\begin{figure*}[ht]
\centering
\includegraphics[width=110mm,clip]{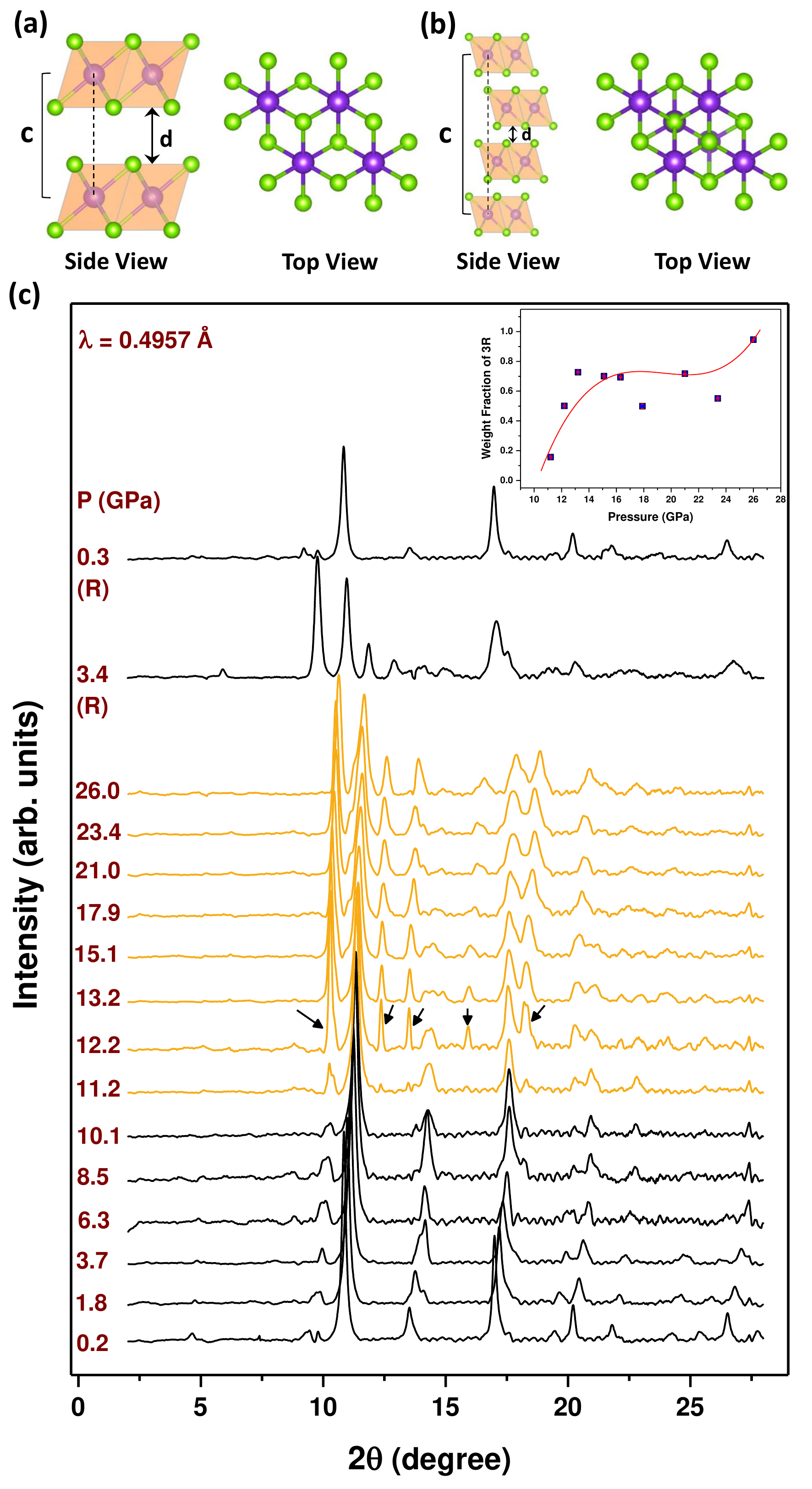}
\caption{\small Side and top view of \textbf{(a)} 1T (\textit{P$\bar{3}$m1}) and \textbf{(b)} 3R (\textit{R$\bar{3}$m}) crystal structures of VSe$_2$. (c) Angle Dispersive XRD patterns during pressurization from 0.2 to 26.0 GPa (two top-most patterns are after depressurizing to 0.3 GPa). Arrows indicate new peaks appearing at the onset of the first order structural transition. The evolution of wight fraction of the 3R phase with increasing pressure is shown in the inset.}
\label{fig1}
\end{figure*}

Angle dispersive powder XRD patterns of VSe$_2$ at room temperature at selected pressure values are stacked in Fig.~\ref{fig1}(c). The ambient trigonal phase with space group \textit{P$\bar{3}$m1}(164), z=1 exhibits stability up to $\sim$10.1 GPa above which new peaks start to appear in the diffraction pattern at $\sim$10.3$\degree$, 12.3$\degree$, 13.5$\degree$, 15.9$\degree$, 18.3$\degree$, 20.3$\degree$, 22.1$\degree$ and 23.8$\degree$ and are prominent in intensity from 12.2 GPa (marked by arrows in Fig.~\ref{fig1}(c)) onwards. Emergence of new Bragg reflections over the existing ones confirms the occurrence of new crystal symmetry coexisting with the previous 1T phase (weight fraction of the 3R phase with increasing pressure is shown in the inset of {Fig.~\ref{fig1}(c)}).

The new high-pressure phase has been indexed unambiguously to a standard CdCl$_2$ rhombohedral 3R structure with a space group symmetry of \textit{R$\bar{3}$m} (166), z=3. This is in contrast to the high-pressure pattern reported by Sereika et al.~\cite{58} showing onset of only small shoulder peaks above 15.5 GPa which were indexed based on \textit{C2/m} (subgroup of \textit{P$\bar{3}$m1}) monoclinic  superstructure. The highest pressure XRD pattern of Fig.~\ref{fig1}(c) confirms that the 1T to 3R transition in VSe$_2$ is incomplete even at our highest measured pressure of $\sim$26 GPa where the two phases still coexist. Also, the structural transition of VSe$_2$ is found to be reversible but with the existence of a large hysteresis of about 8 GPa as indicated by the two topmost patterns of Fig.~\ref{fig1}(c).

\begin{figure*}[ht]
\centering
\includegraphics[width=110mm,clip]{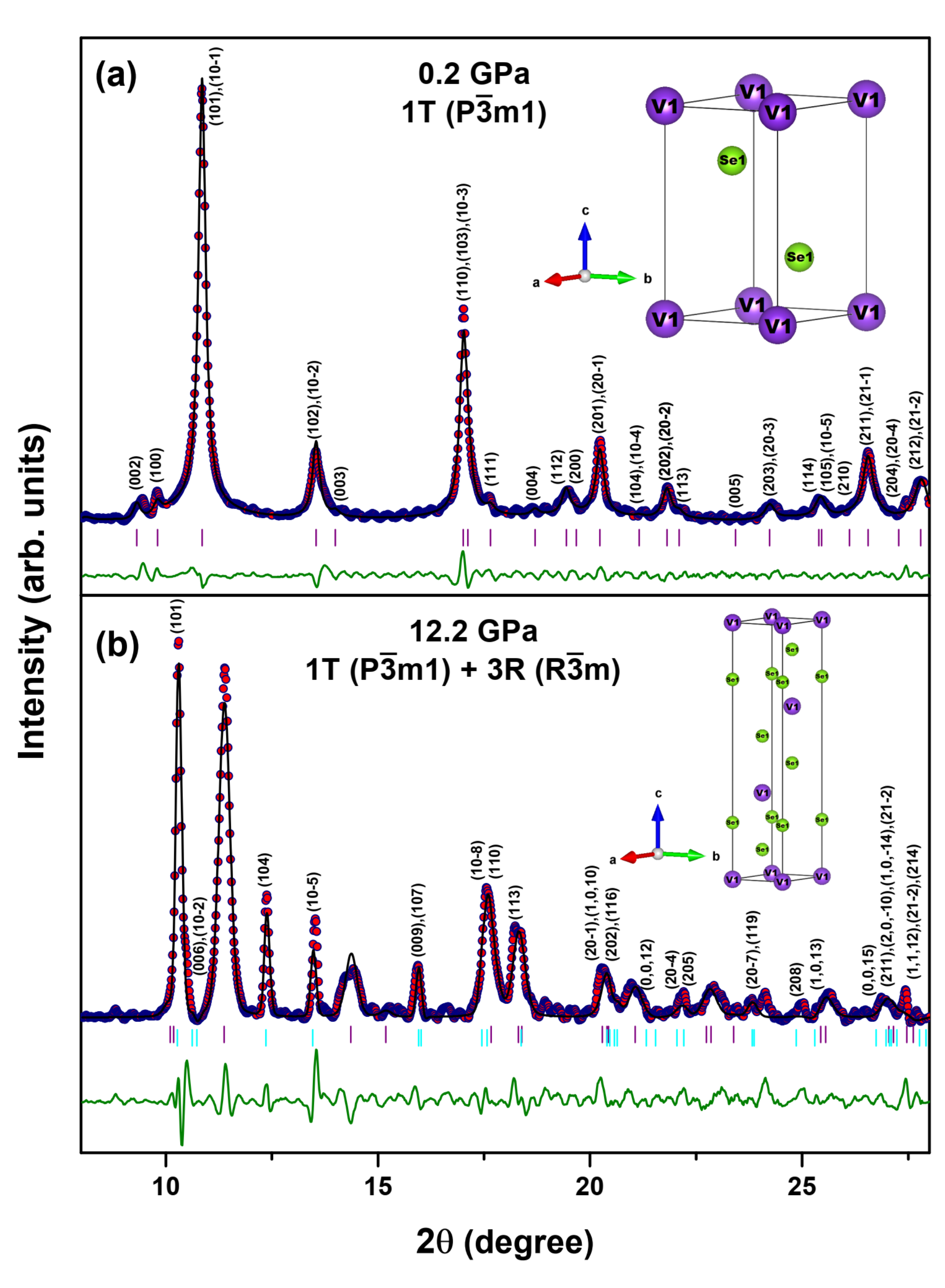}
\caption{\small{\textbf{(a)} and \textbf{(b)} Rietveld refined XRD patterns at 0.2 GPa and 12.2 GPa\rim{,} respectively matched with \textit{P$\bar{3}$m1} ($\#$164) and a mixture of \textit{P$\bar{3}$m1} ($\#$164) and \textit{R$\bar{3}$m} ($\#$166). Experimental data are indicated by solid circles. Calculated pattern is drawn as black solid line. Reflection positions for 1T phase are indicated by magenta vertical bars and those for 3R by cyan ones. Lower dark green curve is the weighted difference between observed and calculated profile. The unit cells including atoms are shown in the inset.}}
\label{fig2}
\end{figure*}

Fig.~\ref{fig2} shows the Rietveld refined fitted patterns at 0.2 GPa and 12.2 GPa using \textit{P$\bar{3}$m1} (1T) and mixture of \textit{P$\bar{3}$m1} (1T) and \textit{R$\bar{3}$m} (3R), respectively. There exist two equivalent representations for the 3R crystal structure \textit{i.e.} using rhombohedral axes leading to the primitive cell (\textit{V}) or by hexagonal axes leading to a unit cell having volume three times (\textit{3V}) with respect to the primitive one. We have adopted the hexagonal axes since it clearly evinces the difference between the stacking sequence of the Se-V-Se tri-layers in the 1T and 3R polytypes as depicted in {Fig.~\ref{fig1}(a) and (b)}. The Rietveld refined parameters including the lattice constants, cell volume, atomic positions of V and Se atoms (in terms of fractional co-ordinates), V-Se bond length, inter-layer distance \textit{d}, and the reduced $\chi^2$ and profile \textit{R$_p$} factors are listed in Table~\ref{table1}.

\begin{figure*}[ht]
\centering
\includegraphics[width=110mm,clip]{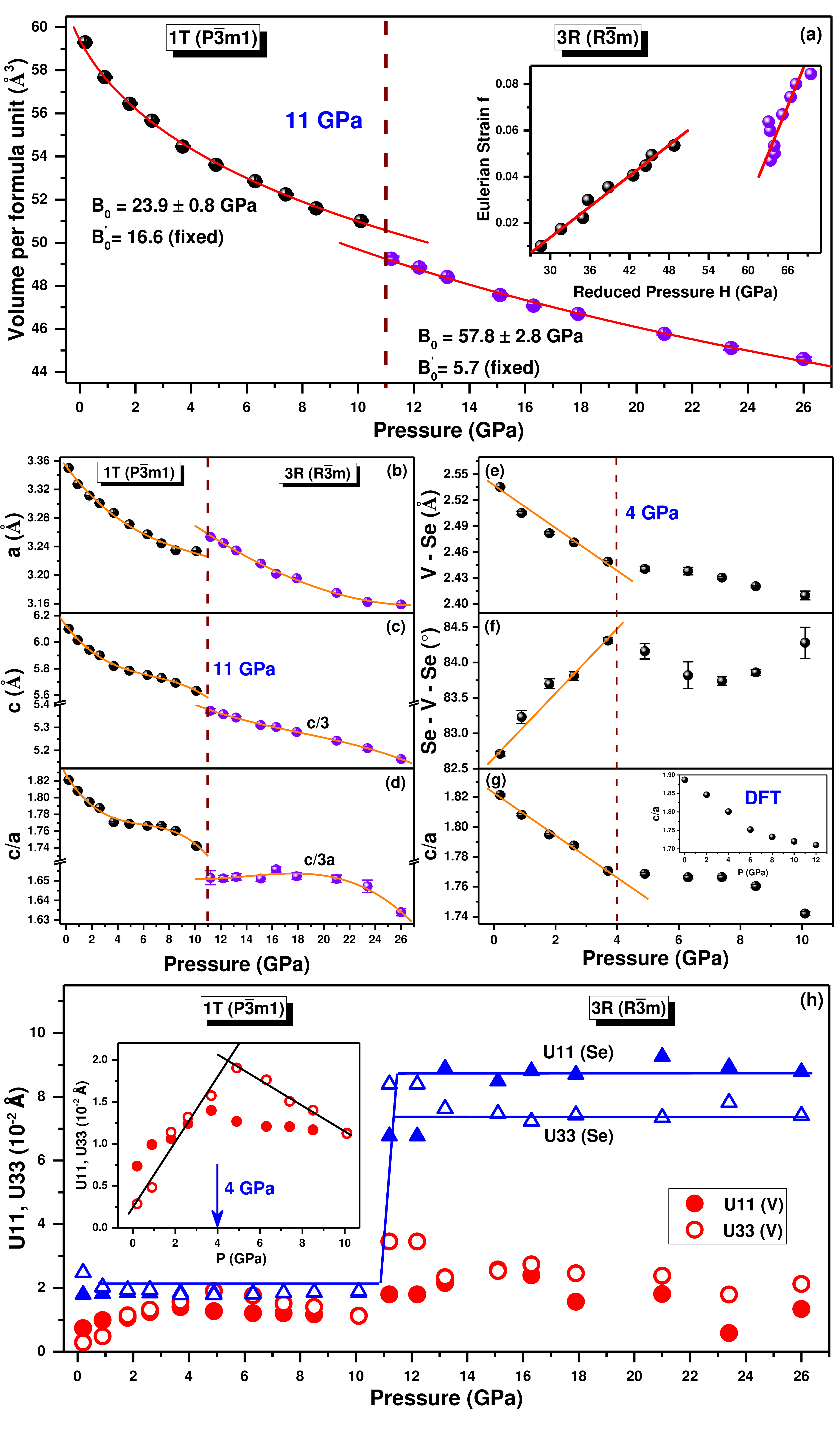}
\caption{\small {\textbf{(a)} Fitted (red solid line) \textit{P-V} diagram using 3$^{rd}$ order BM equation of state. Inset shows Eulerian strain \textit{f} vs reduced pressure \textit{H} plots in the two phases. \textbf{(b)} - \textbf{(d)} Pressure dependence of lattice parameters \textit{a}, \textit{c} and \textit{c/a} ratio for 1T and 3R phases. \textbf{(e)} - \textbf{(g)} Pressure variation of V-Se bond length, Se-V-Se bond angle and \textit{c/a} ratio of 1T phase (the orange solid lines are guide to eye). Inset of \textbf{(g)} shows pressure dependence of \textit{c/a} from DFT calculations. \textbf{(h)} Pressure evolution of the \textit{U11} and \textit{U33} components of the Debye-Waller temperature factors for the V and Se atoms. Inset of \textbf{(h)} shows the subtle anomaly in the \textit{U}-parameters of V atom across the isostructural transition at 4 GPa.}}
\label{fig3}
\end{figure*}

Variation in the lattice parameters \textit{a}, \textit{c} and the \textit{c/a} ratio with increasing pressure (see Fig.~\ref{fig3}(b)-(d)) reveals that the \textit{c/a} ratio of 1T phase decreases up to 10.1 GPa indicating the higher compressibility of \textit{c} axis in contrast to that of \textit{a} (or \textit{b}) axis, giving prominent anisotropy of the crystal which can be attributed to relative strengths of weak inter-layer van der Waals bond and strong intra-layer covalent bonds. In contrast, the \textit{c/a} ratio becomes almost constant after 11 GPa in the 3R phase (excluding highest pressure value). This can be explained from the inter-layer \textit{d} spacing and the V-Se bond length of 3R phase compared to those of 1T at 12.2 GPa (Table~\ref{table1}). A reduced inter-layer separation and a higher V-Se bond length of the 3R phase tunes the interplay between the intra-layer covalent and inter-layer van der Waals couplings resulting in a suppression of the 2D character of the system making it more isotropic and thus giving rise to an almost flat \textit{c/a} ratio with increasing pressure.

The volume per formula unit for each phase determined (see Fig.~\ref{fig3}(a)) against pressure for the entire range to get the \textit{P-V} relation for both the phases. The finite volume discontinuity at $\sim$11 GPa confirms the first order nature of this structural transition which is also corroborated by a large hysteresis ($\sim$8 GPa) in the transition pressure. The \textit{P-V} data in each phase is fitted using third order Birch-Murnaghan (BM) equation of state (EOS)~\cite{3BM} given by,

\scriptsize
\begin{equation}
\label{eqn1}
P=\frac{3}{2}B_0\left[\left(\frac{V0}{V}\right)^{\frac{7}{3}}-\left(\frac{V_0}{V}\right)^{\frac{5}{3}}\right]\lbrace1-\left(3-\frac{3}{4}B_0^{'}\right)\left[\left(\frac{V_0}{V}\right)^{\frac{2}{3}}-1\right]\rbrace
\end{equation}
\normalsize where, \textit{V$_0$} denote the zero-pressure cell volume (per f. u.), \textit{B$_0$} is the zero-pressure bulk modulus and \textit{B$_0^{'}$} is its pressure derivative. In order to get an unambiguous value of these parameters, BM equation can be linearized and cast in terms of reduced pressure $H=\frac{P}{3f\left(1+2f\right)^{5/2}}$, and Eulerian strain $f=\frac{1}{2}\left(X^2-1\right)$, where $X=\left(\frac{V}{V_0}\right)^{-1/3}$ as~\cite{Angel}:
\begin{equation}
\label{eqn2}
H=B_0+\frac{3}{2}B_0\left(B_0^{'}-4\right)f
\end{equation}

The linear fits for the \textit{f} vs \textit{H} plots shown in the inset of Fig.~\ref{fig3}(a) estimate the fitting parameters as, \textit{B$_0$} = 23.8 $\pm$ 0.8 GPa, \textit{B$_0^{'}$} = 16.6 $\pm$ 1.0 in the 1T phase and \textit{B$_0$} = 55.8 $\pm$ 2.0 GPa, \textit{B$_0^{'}$} = 5.7 $\pm$ 0.4 in the 3R phase. These parameters guide us to the fitting of \textit{P-V} data using BM equation of state, giving \textit{V}$_0$ = 59.4 $\pm$ 0.1 $\AA^3$, \textit{B$_0$} = 23.9 $\pm$ 0.8 GPa and \textit{B$_0^{'}$} = 16.6 (fixed) in the 1T phase and \textit{V$_0$} = 56.1 $\pm$ 0.3 $\AA^3$, \textit{B$_0$} = 57.8 $\pm$ 2.8 GPa and \textit{B$_0^{'}$} = 5.7 (fixed) in the 3R phase. We may note that, though the value obtained for \textit{B$_0^{'}$} (= 16.6) in the 1T phase is high as compared to that of other TMDC compounds where it typically ranges from 4 to 11, similar high value of 16.3 for \textit{B$_0^{'}$} was previously reported for pyrite type MnTe$_2$~\cite{FJELLVAG}. We do not detect any change in value of \textit{B$_0$} of the 1T phase around 6 GPa as reported by Sereika et al.~\cite{58} However, as shown in Fig.~\ref{fig3}(e)-(g), pressure dependence of the V-Se bond length, Se-V-Se bond angle and \textit{c/a} ratio of the 1T phase shows a significant change, making an isostructural transition at $\sim$4 GPa, in agreement with the recent reports~\cite{57,58}.

The pressure effects on the Debye-Waller factor (DWF)~\cite{Debye, Waller} \textit{U$_{ij}$}, a measure of the mean-square thermal displacement of an atom from its equilibrium position due to crystal lattice vibrations, has not been studied hitherto fore in diffraction measurements. Figure~\ref{fig3}(h) shows the pressure variation of the diagonal components \textit{U11} (= \textit{U22}) and \textit{U33} of the thermal ellipsoid for the V and Se atoms in 1T and 3R phase of VSe$_2$. The \textit{U}-parameters for the V atom in the 1T phase show subtle anomalous behaviour (inset of Fig.~\ref{fig3}(h)) across the isostructural transition pressure of $\sim$4 GPa. In the 3R phase, the \textit{U}-parameters for the V atoms do not show any significant variation. Notably, in contrast, for the Se atoms, both \textit{U}-parameters \textit{U11} and \textit{U33} remain almost constant in the 1T phase but show an abrupt increase ($\sim$4 times) in the 3R phase. In this regard, we may note that the abrupt change in the \textit{U}-parameters across the 11 GPa transition cannot be taken care of by the phase fraction of 1T and 3R phases (see SI~\cite{suppl} section III).

\begin{figure*}[ht]
\centering
\includegraphics[width=140mm,clip]{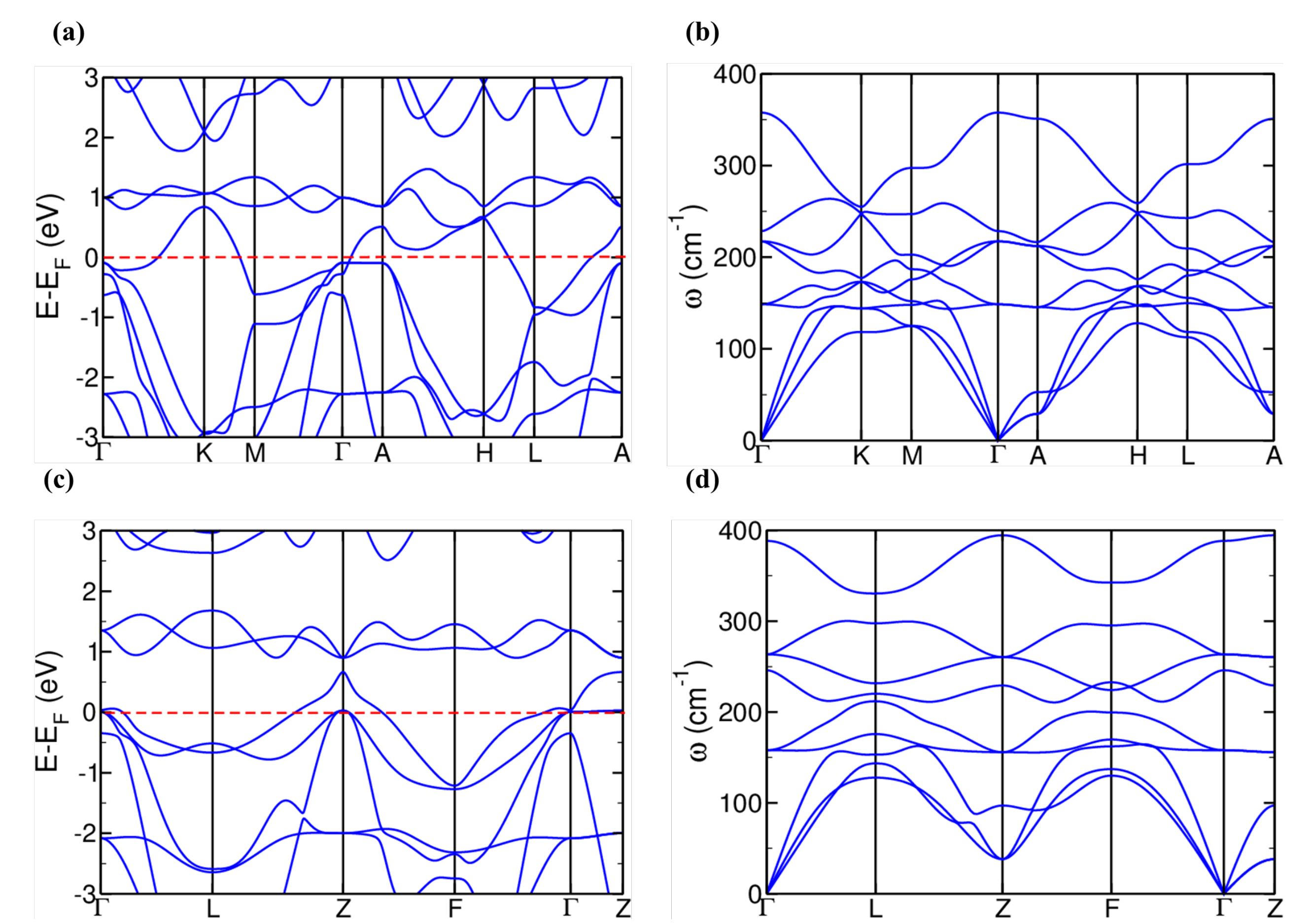}
\caption{\small {\textbf{(a)} Electronic structure and \textbf{(b)} phonon dispersion of 1T-VSe$_2$ (space group: \textit{P$\bar{3}$m1}) at 0 GPa, \textbf{(c)}  Electronic structure and \textbf{(d)} phonon dispersion of 3R-VSe$_2$ (space group: \textit{R$\bar{3}$m}) at 12 GPa.}}
\label{fig4}
\end{figure*}

We now present our first-principles computational methods to understand the structural evolution in VSe$_2$. Our calculations reproduce the metallic nature of ambient 1T-VSe$_2$ (\textit{P$\bar{3}$m1}) as shown in Fig.~\ref{fig4}(a). Our estimates of the optimized lattice constants of 1T-VSe\textsubscript{2} at 0 GPa are \textit{a} = 3.35 {\AA} and \textit{c} = 6.12 {\AA}, in close agreement with our experimental values (\textit{a} = 3.35 {\AA} and \textit{c} = 6.10 {\AA}). The phonon-dispersion (Fig.~\ref{fig4}(b)) of 1T-VSe$_2$ at 0 GPa confirms its local stability as no imaginary frequencies were observed. 3R-VSe$_2$ (\textit{R$\bar{3}$m}) is metallic at 12 GPa since the valence band maxima and conduction band minima overlap (Fig.~\ref{fig4}(c)). Optimized lattice parameters obtained from our first-principles calculation using Grimme-D2 van der Waals correction at 12 GPa are \textit{a} = 3.16 {\AA} and \textit{c} = 17.39 {\AA}, where \textit{c}-value is overestimated as compared with the experimental values (\textit{a} = 3.24 {\AA} and \textit{c} = 16.07 {\AA}). The \textit{R$\bar{3}$m} phase is dynamically stable with no unstable phonon modes in its phonon dispersion (Fig.~\ref{fig4}(d)). Lattice parameters of 1T-VSe$_2$ vary smoothly as a function of hydrostatic pressure up to \textit{P} $\sim$12 GPa, with a notable change in the slope of \textit{c/a} ratio with pressure at \textit{P$_C$} $\sim$6 GPa, (inset of Fig.~\ref{fig3}(g)) suggesting an isostructural phase transition, consistent with our experimental results and the recent report~\cite{58}.

To explore the possibility of a pressure dependent phase transition from 1T to 3R structure of VSe$_2$, we have determined the changes in enthalpy of these structures but did not observe any transition from the 1T to 3R phase (see SI~\cite{suppl} Fig. \textcolor{blue}{S2}). VSe$_2$ is a layered material which has a strong covalent bonding within the layer and weak van der Waals interaction between the layers. In this regard, we compared the lattice parameters of 3R with experiments obtained using different flavours of van der Waals correction, London-s6 forces and introduced Hubbard U parameter of 1 eV~\cite{Dudarev} to include on-site correlations of \textit{d} electrons of the V atom. These results are presented in SI~\cite{suppl} Table \textcolor{blue}{S2}. As we did not find a phase transition from 1T to 3R structure under hydrostatic pressure, we investigated the stability of 2H$_a$ (another polytype into which bulk di-chalcogenides crystallize) having a hexagonal unit cell with \textit{aBa cBc} stacking and space group \textit{P6$_3$/mmc}~\cite{esters2017dynamic,61}. Phonon dispersion at 12 GPa confirms that 2H\textsubscript{a} is stable and has soft modes indicating a possible phase transition. From estimated difference in enthalpy, we do find a transition from 1T to 2H\textsubscript{a} near \textit{P} $\sim$12 GPa. Though our theory predicts this phase transition, the 2H$_a$ structure cannot be fitted to our XRD data at high pressures. We have also considered 3R structure based on 2H$_a$ stacking (\textit{R3m} space group with \textit{aBa bCb cAc} stacking). The relative stability of this 3R with respect to 1T structure as seen from difference in enthalpy does not reveal a phase transition (see SI~\cite{suppl} section V for more information on stability analysis of 2H$_a$ and 3R (R3m)).

To examine if the finite-temperature effects contribute to the stability of the 3R phase, we have evaluated temperature dependent vibrational free energies of 1T, 3R (\textit{R$\bar{3}$m}) and 3R (\textit{R3m}) structures. As evident in the transition temperatures at various hydrostatic pressure (see SI~\cite{suppl} section VI for details), the temperature that stabilizes 3R polytypes is not realistic, ruling out temperature effects.

Having examined all the above possibilities to stabilize the 3R phase established unambiguously in our XRD experiments, we obtain the energetics of 1T and 3R structures of VSe$_2$ with spin-polarized calculations using VASP including Hubbard parameter correction GGA+\textit{U}, (\textit{U}\textsubscript{eff} = \textit{U} - \textit{J} = 8.0 eV). Sampling of Brillouin zone (BZ) integrations was carried out on dense uniform 24×24×12 and 24×24×24 meshes of k-points for 1T and 3R structures of VSe$_2$, respectively. Calculated enthalpies of 1T and 3R phases as a function of pressure reveal a phase transition from 1T to 3R structure at \textit{P} $\sim$9 GPa.

\section{\bf{CONCLUSIONS}}\vspace{0.2cm}

To conclude, we address two pressure driven transitions in bulk 1T-VSe$_2$ using X-ray diffraction and DFT studies. The first transition around 4 GPa is isostructural with distinctive anomalies in bond length, bond angle, \textit{c/a} ratio and the Debye-Waller factors. The second transition around 11 GPa is from 1T (\textit{P$\bar{3}$m1}) to 3R (\textit{R$\bar{3}$m}) structure due to the sliding of Se-V-Se tri-layer leading to a contraction of unit cell volume per formula unit by $\sim$3$\%$. Similar layer sliding mechanism has also been seen in other TMD materials like MoS$_2$,~\cite{49,50} and WSe$_2$~\cite{52}. The 1T to 3R transition witnesses a relatively large jump in the thermal factors for the selenium atoms which can be related to both the disorder and the enhanced anharmonic interactions in the high-pressure phase. However, presently we are not aware of any mechanism for enhanced disorder across the transition, and this needs further investigation in future. Our analysis based on first-principles calculations could confirm stability of 3R phase around 9 GPa only after incorporation of spin-polarized calculations to account for the Hubbard correction with \textit{U$_{eff}$} = \textit{U} - \textit{J} = 8 eV. Our DFT calculations also exclude any possibility of transition to a monoclinic superstructure phase above 15.5 GPa in the system (see SI~\cite{suppl} section VII for more information).

\section{\bf{ACKNOWLEDGEMENTS}}\vspace{0.2cm}

AKS thanks Nanomission Council and the Year of Science professorship of DST for financial support. LH acknowledges the financial support from the Department of Science and Technology (DST), India [Grant No. SR/WOS-A/PM-33/2018 (G)], IISER Pune for providing the facilities for crystal growth and characterization. We thank Boby Joseph for his support during XRD measurements at the Xpress beamline of Elettra Sincrotrone Trieste. Financial supports by the Department of Science and Technology (DST) of the Government of India is also gratefully acknowledged. KD is thankful to Jawaharlal Nehru Centre for Advanced Scientific Research, India, for a research fellowship. UVW is grateful to SERB-DST for support through a JC Bose National fellowship.

\bibliographystyle{ieeetr}

\newpage

\section{Tables and Figures}

\begin{figure*}[h!]
\centering
\includegraphics[width=140mm,clip]{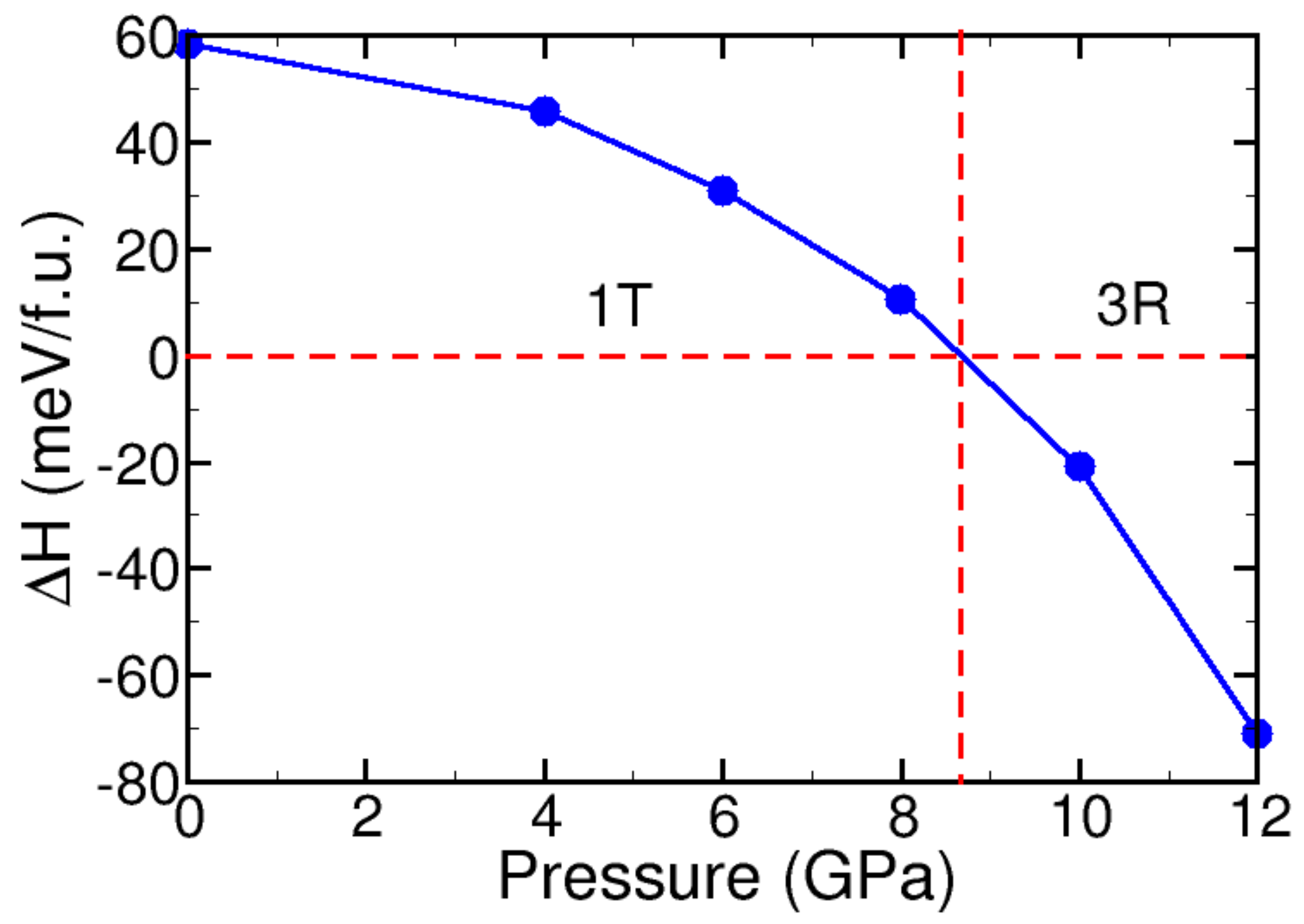}
\caption{\small{The difference in enthalpy between 3R (space group: \textit{R$\bar{3}$m}) and 1T structures of VSe$_2$, shows a phase transition from 1T to 3R structure of VSe$_2$ at \textit{P} $\sim$9 GPa.}} 
\label{fig5}
\end{figure*}

\begin{table*}[ht]
\caption{\footnotesize Rietveld refined parameters}
    \begin{center}
    \begin{tabularx}{\columnwidth}{X X X X X}
    \arrayrulecolor{black}\hline \hline \\ [-1.5ex]
    \multicolumn{1}{c}{}
    & \multicolumn{1}{c}{\textbf{0.2 GPa (1T)}}
    & \multicolumn{1}{c}{\textbf{12.2 GPa (1T)}}
    & \multicolumn{1}{c}{\textbf{12.2 GPa (3R)}} \\[1ex] \hline\\ 
    \textbf{Space Group} & \multicolumn{1}{c}{\textit{P$\bar{3}$m1}(z=1)} & \multicolumn{1}{c}{\textit{P$\bar{3}$m1}(z=1)} & \multicolumn{1}{c}{\textit{R$\bar{3}$m}(z=3)}\\[1ex]
    \textbf{\textit{a} (\AA)}   & \multicolumn{1}{c}{$3.349998$} & \multicolumn{1}{c}{$3.226370$} & \multicolumn{1}{c}{$3.244756$}\\[1ex]
    \textbf{\textit{c} (\AA)}   & \multicolumn{1}{c}{$6.101036$} & \multicolumn{1}{c}{$5.628404$} & \multicolumn{1}{c}{$16.073330$}\\[1ex]
    \textbf{\textit{V}/f.u. (\AA$^3$)} & \multicolumn{1}{c}{$59.296$} & \multicolumn{1}{c}{$50.739$} & \multicolumn{1}{c}{$48.852$}\\[1ex]
    \textbf{V (X,Y,Z)} & \multicolumn{1}{c}{(0,0,0)} & \multicolumn{1}{c}{(0,0,0)} & \multicolumn{1}{c}{(0,0,0)}\\[1ex]
    \textbf{Se (X,Y,Z)} & \multicolumn{1}{c}{$(\frac{1}{3},\frac{2}{3},0.268636)$} & \multicolumn{1}{c}{$(\frac{1}{3},\frac{2}{3},0.268636)$} & \multicolumn{1}{c}{$(0,0,0.221085)$}\\[1ex]
    \textbf{V-Se (\AA)} & \multicolumn{1}{c}{$2.5352$} & \multicolumn{1}{c}{$2.3863$} & \multicolumn{1}{c}{$2.6009$}\\[1ex]
    \textbf{\textit{d} (\AA)} & \multicolumn{1}{c}{$2.823120$} & \multicolumn{1}{c}{$2.645507$} & \multicolumn{1}{c}{$1.749368$}\\[1ex]
    \textbf{Reduced} \boldsymbol{$\chi^2$} & \multicolumn{1}{c}{$5.31$} & \multicolumn{2}{c}{$8.50$}\\[1ex]
    \textbf{Profile \textit{R$_p$}} & \multicolumn{1}{c}{$14.09\%$} & \multicolumn{2}{c}{$21.22\%$}\\ [1ex]
    \hline
    \end{tabularx}
    \label{table1}
    \end{center}
\end{table*}

\end{document}